\documentclass[letterpaper, 10 pt, conference]{ieeeconf}  

\IEEEoverridecommandlockouts                              
\usepackage[utf8]{inputenc}
\usepackage[T1]{fontenc}
\usepackage{acronym}
\usepackage{graphicx}
\usepackage{multirow}
\usepackage{microtype}
\usepackage{makecell}
\usepackage{subcaption,siunitx,booktabs} 
\usepackage{authblk}
\usepackage{subcaption}
\usepackage{caption}
\usepackage{times}
\usepackage{epsfig}
\usepackage{graphicx}
\usepackage{amsmath}
\usepackage{amssymb}
\usepackage{acronym}
\usepackage{microtype}
\usepackage[english]{babel}

\title{\LARGE \bf
Structure Guided Manifolds for Discovery of Disease Characteristics}

\author{Siyu Liu$^{1}$, Linfeng Liu$^{1}$, Xuan Vinh$^{2}$, Stuart Crozier$^{1}$, Craig Engstrom$^{1}$, Fatima Nasrallah$^{2}$, Shekhar Chandra}

\affil[$^{1}$]{School of Information Technology and Electrical Engineering, University of Queensland, Australia}
\affil[$^{2}$]{Queensland Brain Institute, University of Queensland, Australia}
\begin{document}
\maketitle
\thispagestyle{empty}
\pagestyle{empty}
\begin{abstract}
In medical image analysis, the subtle visual characteristics of many diseases are challenging to discern, particularly due to the lack of paired data. For example, in mild \ac{AD}, brain tissue atrophy can be difficult to observe from pure imaging data, especially without paired \ac{AD} and \ac{CN} data for comparison. This work presents \ac{DiDiGAN}, a weakly-supervised style-based framework for discovering and visualising subtle disease features. DiDiGAN learns a disease manifold of AD and CN visual characteristics, and the style codes sampled from this manifold are imposed onto an anatomical structural ``blueprint" to synthesise paired \ac{AD} and \ac{CN} \acp{MRI}. To suppress non-disease related variations between the generated \ac{AD} and \ac{CN} pairs, \ac{DiDiGAN} leverages a structural constraint with cycle consistency and anti-aliasing to enforce anatomical correspondence. When tested on the \ac{ADNI} dataset, \ac{DiDiGAN} showed key \ac{AD} characteristics (reduced hippocampal volume, ventricular enlargement, and atrophy of cortical structures) through synthesising paired \ac{AD} and \ac{CN} scans. The qualitative results were backed up by automated brain volume analysis, where systematic pair-wise reductions in brain tissue structures were also measured.
\end{abstract}

\acresetall
\section{Introduction}
Discovering the visual characteristics of disease states from medical imaging data can provide important insights into the presence, progression and effects of pathological changes. 
In many brain pathologies, magnetic resonance (MR) images provide excellent visual characterization of lesions~\cite{brats_dataset} although visualizing very early or subtle changes in brain volume in conditions such as \ac{AD} can be extremely challenging. Subtle AD features are difficult to discern and visualise due to the lack of paired \ac{AD} and \ac{CN} data to provide anatomical correspondence. Hence, experts need to know what to specifically look for. With differentiable vision systems like \acp{CNN}, it is possible to automatically extract \textit{some} brain features with methods like GradCAM~\cite{gradcam}. However, GradCAM does not communicate any specific visual characteristics beyond rough locations of abnormalities. In many cases, the specific effects of a disease still require side-by-side comparisons to observe.

In theory, \acp{GAN}~\cite{gan} are capable of creating synthetic medical images of an anatomical structure in different disease conditions. Style-based~\cite{stylegan1}\cite{stylegan2} \acp{GAN} are of particular interest for such tasks as they provide mechanisms for maintaining anatomical correspondence. The generator learns a high-dimensional latent space where style codes are sampled and then injected into the generator to manipulate the output image. It was shown that the learnt latent space forms a smooth high-dimensional manifold that facilitates image interpolation~\cite{stylegan1}. Hence, style-based \acp{GAN} have been linked toc manifold and latent representation learning \cite{wu2021stylespace}, \cite{patashnik2021styleclip}.
Subsequent works, including StarGANs~\cite{stargan1}\cite{stargan2} demonstrated unpaired image translation using StyleGAN. 
It was shown that the learnt manifold (style) could be separated to govern only a select subset of features, leaving the rest of the content intact. This idea has also been demonstrated for medical image analysis~\cite{distentangle}, where (medical imaging) modality information was successfully separated from anatomical information. 
These works suggest style-based \acp{GAN} could indeed learn a manifold of \ac{AD} features to synthesise paired \ac{AD} and \ac{CN} scans. However, as will be shown in this paper, simply applying these methods does not produce usable synthetic \ac{AD} and \ac{CN} samples with sufficient anatomical correspondence (\ac{AD} features, if any, are concealed by non-\ac{AD} structural variations between the sample pairs).

The inability of style-based \acp{GAN} to maintain anatomical correspondence can be attributed to aliasing, which is a substantive problem often overlooked in deep learning. Recently, it was reported that anti-aliasing could improve the accuracy and generalisation of \acp{CNN}~\cite{aa} for classification. There have been similar findings in generative modelling. The highlight of StyleGAN3~\cite{stylegan3} was the introduction of anti-aliasing, which mitigates ``texture sticking" artifacts during latent space interpolation. While these artifacts do not significantly deteriorate image quality, they invalidate applications requiring smooth manifold interpolation. Once anti-aliasing was implemented following the Shannon-Nyquist sampling theorem~\cite{Shannon}, traversing the manifold of StyleGAN3 yields more natural transitions in the image space. This work recognises the unique aspects of anti-aliasing for establishing a reliable pair-wise anatomical correspondence. Even small artifacts in the feature maps can morph into significantly dissimilar anatomical structures. Therefore the aliasing issue must be addressed.

This current work presents \ac{DiDiGAN}, a style-based network capable of synthesising paired \ac{AD} and \ac{CN} MR images while maintaining anatomical correspondence. Many key \ac{AD} features are made observable even to the untrained eye by simple visual comparison of the generated image pairs. The highlights of this network include

\begin{itemize}
    \item A learnt disease manifold that encodes \ac{AD} and \ac{CN} characteristics. The manifold is then injected into the generator to control the expression of \ac{AD} features (or the lack thereof) in the output.
    
    \item The generative process is guided by a structural constraint. The constraint is low-resolution by design to establish high-level anatomical correspondence between the generated \ac{AD} and \ac{CN} image pairs. Its lack of fine details also leaves sufficient room for \ac{AD} characteristics (style) to be synthesised. The constraint is enforced using cycle consistency.
    
    \item The first to employ anti-aliasing to enforce structural correspondence (further to the above constraint) between generated image pairs.
    
    \item The disease manifold is automatically built in a weakly supervised manner where only disease class labels are required.

\end{itemize}

\ac{DiDiGAN} was tested on the \ac{ADNI} dataset, where it synthesised paired \ac{AD} and \ac{CN} MR images. The generated image pairs clearly exhibit key \ac{AD} characteristics including, reduced hippocampal volume, ventricular enlargement and cortical atrophy while maintaining pair-wise anatomical correspondence. The results were first examined by three clinical experts followed by Jacobian index analysis and brain volume analysis where systematic brain shrinkage was measured.

\section{Methods}
The idea formulation of \ac{DiDiGAN} revolves around projecting a learnt disease manifold onto a shared anatomical ``blueprint" to synthesise comparable \ac{AD} and \ac{CN} images. This section describes the components of \ac{DiDiGAN} and the mechanisms employed to synthesise disease characteristics without compromising anatomical correspondence.

\begin{figure*}[ht]
 \centering
 \includegraphics[width=1\linewidth]{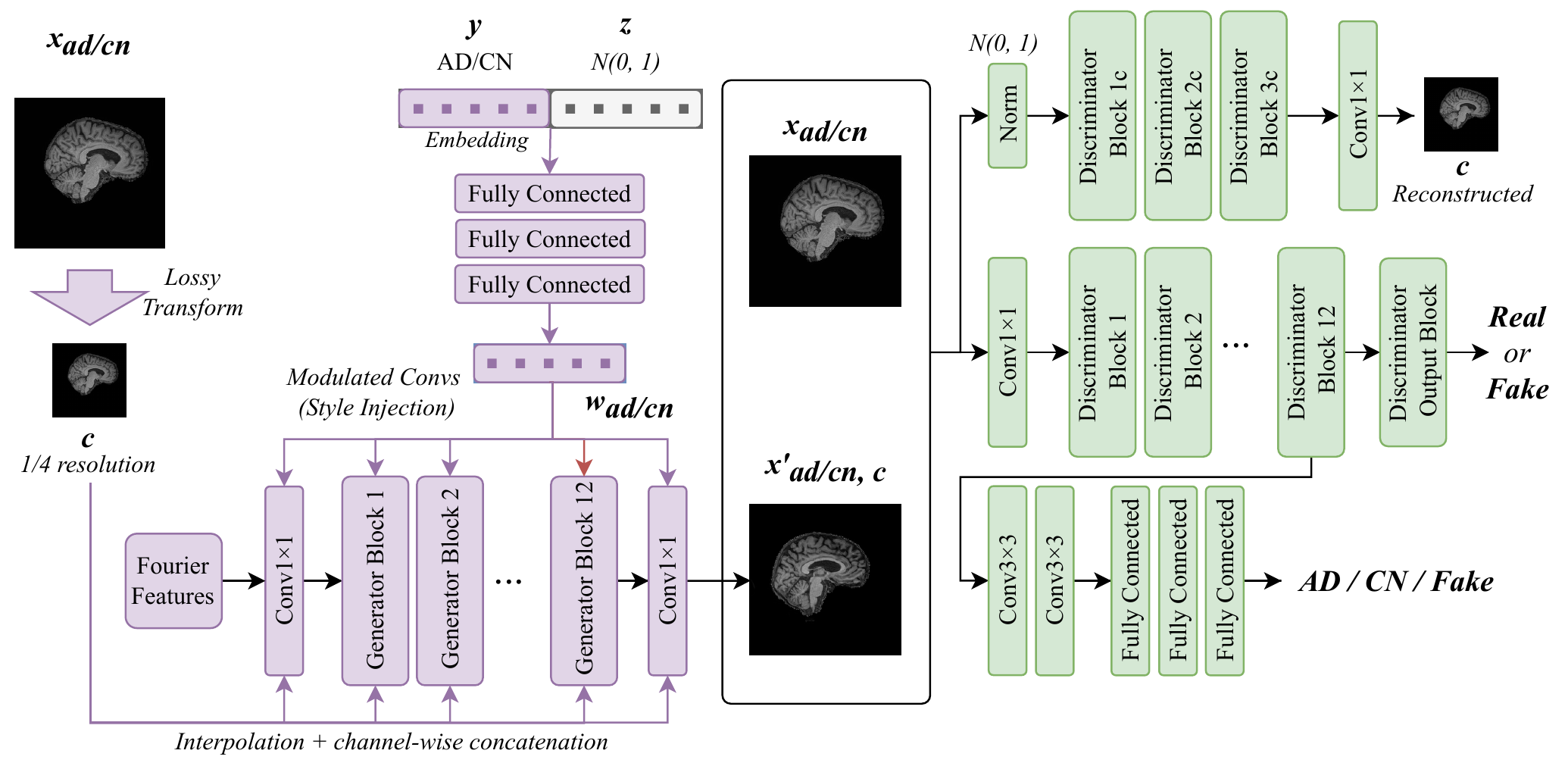}
 \caption{Architecture of \ac{DiDiGAN}. The alias-free generator is conditioned on both the constraint $c$ and the disease class label $y$. The discriminator attempts to recover the input $c$ and $y$ through adversarial training.}
 \label{fig:architecture}
\end{figure*}

The architecture of \ac{DiDiGAN} is shown in Figure~\ref{fig:architecture}. The procedure to generate a pair of contrasting samples $x_c' \in \{x'_{ad,c}, x'_{cn,c}\}$ is described by $x_{ad,c}'=G(w_{ad},c,z),\ x_{cn,c}'=G(w_{cn},c,z)$, where
\begin{itemize}
    \item $w \in \{w_{ad}, w_{cn}\}$ are 512-d disease style codes sampled from the learnt disease manifold, in this case, AD or the lack thereof (CN). $w_{ad}$ and $w_{cn}$ are conditioned on the embeddings of the class labels $y_{ad}$ and $y_{cn}$, respectively. Each embedding is concatenated with a noise vector $z$ (for diverse outputs) and mapped to $w$ using three fully connected layers of 512 units. $w$ is injected into $G$ as style such that the generated images can reflect \ac{AD} characteristics.
    
    \item $c$ is a deliberately lossy representation of an input image $x$ which acts as a rough ``blueprint" for $x_{ad,c}'$ and $x_{cn,c}'$ to be structurally similar at a high-level. Its vagueness deliberately leaves room for specific disease characteristics (AD or CN) to be incorporated through style injection. An overly precise constraint would exert too much control over the content, diminishing disease style codes' effect. In the present work, $c$ is simply a low-resolution version of $x$ (scaled with anti-aliasing). The scaling factor is a hyper-parameter controlling the amount of room left for \ac{AD} features to develop. It was also found that other lossy representations such as segmentation maps and edge maps would also work as alternative constraints.
    
    \item $G$ is an alias-free convolutional generator with a starting resolution of $4\times4$ and an output resolution of $256\times256$. There are 12 convolutional blocks that gradually up-samples the image. $w$ is incorporated via modulated convolution and $c$ is imposed by channel-wise concatenation (after re-scaling as needed). While the constraint $c$ establishes high-level anatomical correspondence, anti-aliasing plays a crucial role in suppressing more fine-grained undesirable variations between $x'_{ad,c}$ and $x'_{cn,c}$ which are not controlled by $c$. All the operations in the generator are made alias-free following Shannon-Nyquist sample theorem, and the anti-aliasing procedures involve (or approximate) I) casting the discrete signals to the continuous domain followed by II) low-pass filtering and then III) discrete sampling back to discrete domain~\cite{stylegan3}. The key here is all the undesirable high-frequency signals in the feature space are eliminated to ensure no disruptions to anatomical correspondence.
    
\end{itemize}

The discriminator is a \ac{CNN} with three prediction heads $D_{adv}$, $D_{C}$ and $D_{W}$, which predict I) the logit for realness, II) the reconstructed structural constraint and III) the classification of disease, non-disease or fake, respectively. These prediction heads correspond to three optimisation objectives $L_{adv}$, $L_{C}$ $L_{W}$.

\begin{gather*}
    \min _{{D}}\max _{{G}} L_{adv} = \mathbb{E}[\textnormal{log}D_{adv}(x)]~-~\mathbb{E}[\textnormal{log}(D_{adv}(x'_c))]\\
    \min _{{D}}\min _{{G}} L_{C} = ||c - D_C(x'_c)||^2\\ 
    \min _{{D}}\max _{{G}} L_{W} = -\sum_{i \in Y} y_{i} \dot \log{(D_W(X))}, \\\textnormal{where}\ X \in \{x'_{c}, x\}, \ Y \in \{ad, cn, fake\}
\end{gather*}

$L_{adv}$ is the main \ac{GAN} loss which is non-saturating for improved learning signals. $L_{C}$ is an mean squared error (MSE) cycle consistency loss minimised by $G$ and $D$ to supervise $X_c'$'s visual adherence to $c$. Since $c$ is low-resolution by design, $L_{C}$ would only enforce high-resolution (``vague") anatomical correspondence for $G$. This is where anti-aliasing functions to enforce anatomical correspondence at finer scales. $L_{W}$ is the standard cross-entropy loss for classifying $x'_c$ and $x$ into $ad$, $cn$ or $fake$ ($fake$ is an adversarial class for $D_{W}$ to learn classification through adversarial training). From $G$'s perspective, $L_{W}$ supervises the incorporation of disease style codes such that, for example, $D_{W}$ would predict $ad$ for $X_{ad, c}'$. There are also other regularisation loss functions including the path-length regularisation $L_{pl}$~\cite{stylegan3}, $R1$ gradient penalty $L_{R1}$~\cite{gan_convergence} and an explicit diversification loss $L_{div} = -||G(w,c,z_1) - G(w,c,z_2)||$ to ensure the manifold is well-formed. The total loss $L_{total}$ is then defined as
\begin{gather*}
    L_{total} = L_{adv} + L_{C} + L_{W} + L_{div} + L_{pl} + L_{R1}
\end{gather*}

\subsubsection{Data Preparation}
The dataset analysed in the present work consists of T1-weighted MRI images from the ADNI2 and ADNI3 datasets. There are 1565 3D MR imaging examinations from 942 patients, and the scans were divided into training, validation and test sets following patient-level split with a ratio of 0.6, 0.2 and 0.2. Standard prepossessing procedures (N4 bias-field correction~\cite{n4}, skull stripping using SPM \cite{tzourio2002automated}) were applied to the entire dataset.

\section{Experiments and Results}
The disease characteristic modelling and visualisation was performed by generating contrasting coronal \ac{AD} and \ac{CN} slices. The same approach was also performed on the sagittal slices to verify the coronal view findings. For each 3D brain scan, the centre 40 slices were extracted and zero-padded to $256 \times 256$ pixels. Down-sampled versions of these slices ($64 \times 64$) were used for $c$ in the process to generate contrasting samples. All the experiments involving \ac{DiDiGAN} are completed using a Tesla V100 GPU, and training takes three days. The generated synthetic images pairs were examined by three clinical experts followed by qualitative and quantitative analyses. 
\begin{figure*}
     \centering
     \begin{subfigure}[b]{1\textwidth}
              \centering
         \includegraphics[width=0.9\textwidth]{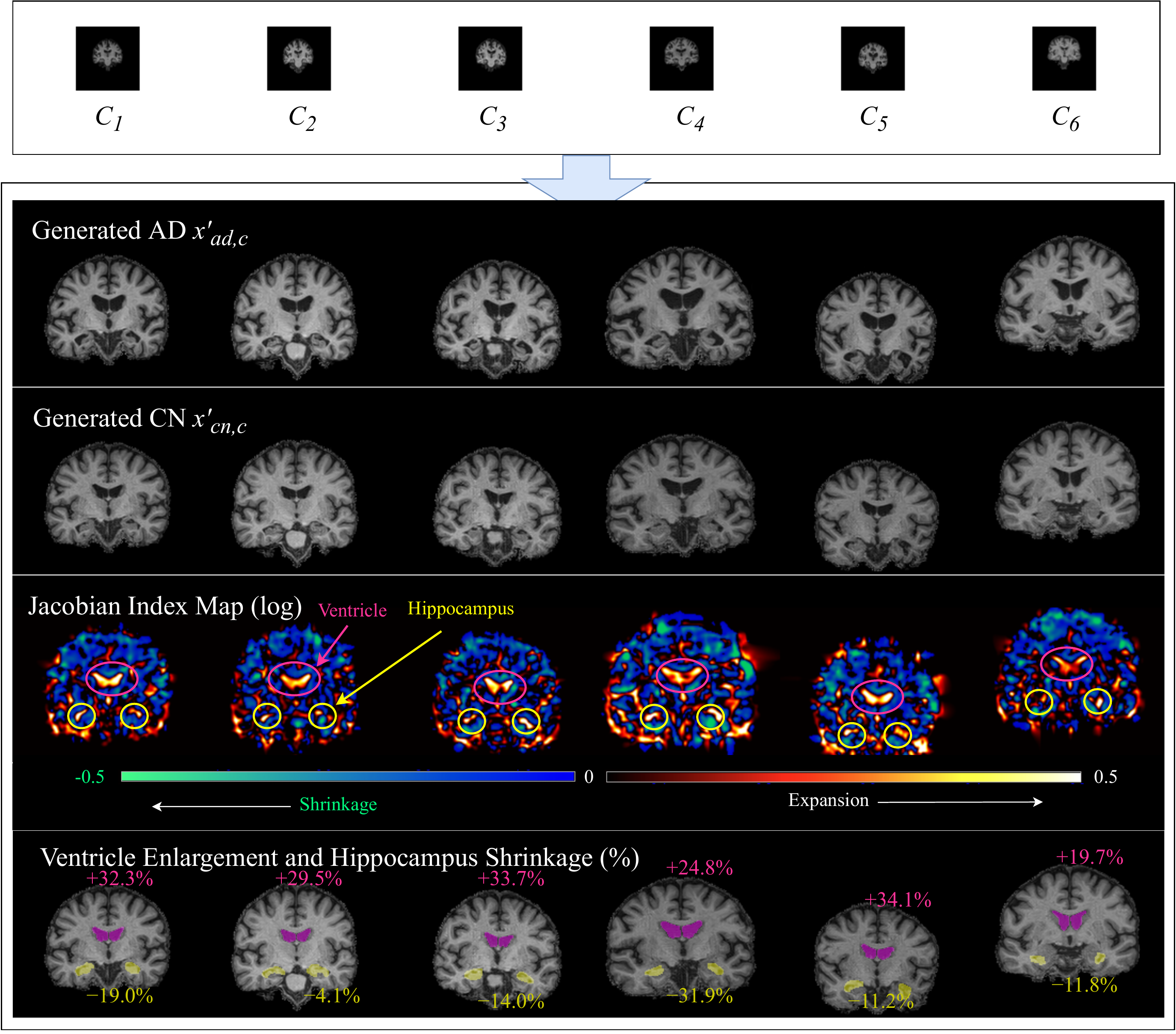}
         \caption{Examples of generated \ac{AD} $x'_{ad,c}$, \ac{CN} $x'_{cn,c}$ coronal slices and Jacobian index maps in log scale. The Jacobian index values are associated with degrees of brain tissue expansion and shrinkage. Ventricular expansion and reduced hippocampal volume were estimated based on Jacobian index values of the manually drawn regions of interest (ROIs) (bottom sub-figure). Steps to acquire the Jacobian index maps and shrinkage estimates can be found in section S1.} 
        \label{fig:jac}
     \end{subfigure}
        
     \begin{subfigure}[b]{1\textwidth}
        \centering
         \includegraphics[width=0.9\textwidth]{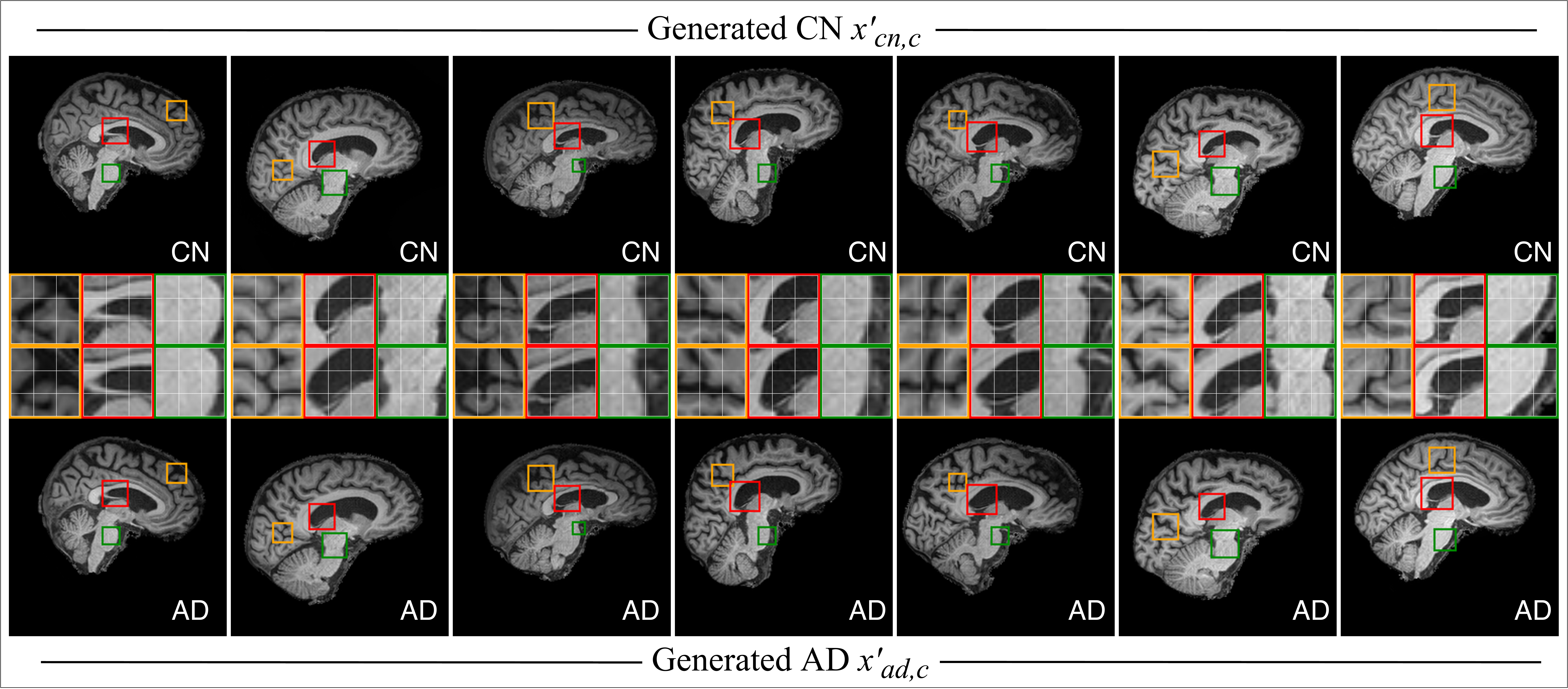}
         \caption{Examples of generated \ac{AD} $x'_{ad,c}$, \ac{CN} $x'_{cn,c}$ sagittal slices also showing ventricular enlargement and cortical atrophy.}
        \label{fig:sag}
     \end{subfigure}

        \caption{Example synthesised \ac{AD} and \ac{CN} pairs from \ac{DiDiGAN}  (more examples in Fig.~\ref{fig:s4}). 2D Jacobian index was computed for the coronal view.}
        \label{fig:three graphs}
\end{figure*}

Figure~\ref{fig:jac} shows examples of synthesised coronal \ac{AD} ($x_{ad,c}'$) and \ac{CN} ($x_{cn,c}'$) image pairs demonstrating high degrees of pair-wise structural similarities indicating good anatomical correspondence.
\ac{AD}-associated brain volume loss between each sample pair is quantified by the corresponding Jacobian index map~\cite{ants} (registration-based morphometry). Consistent with the expected progression of \ac{AD}\cite{schuff2009mri}\cite{sabuncu2011dynamics}, ventricular enlargement, reduced hippocampal volume and cortical atrophy are evident (high Jacobian index values indicating regional expansion in \ac{CSF}). Figure~\ref{fig:jac} (bottom) shows manually drawn regions of interest for the ventricular and hippocampal areas. Across the 6 sample pairs, the Jacobian index values suggest an average ventricular area increase of 29.0\%  and an average hippocampal area reduction of 15.3\% reduction. The coronal view results are further corroborated by an independent experiment performed on the sagittal view slices where reduced hippocampal volume and cortical atrophy are also visible through side-by-side comparison (Figure~\ref{fig:sag}). A StarGANv2 and a CycleGAN were also trained as baseline methods, but no usable results could be obtained for comparative analysis. StarGAN failed to establish anatomical correspondence, while CycleGAN failed to convey any difference between \ac{AD} and \ac{CN}, at all (Fig.~\ref{fig:s1} and \ref{fig:s2}).

Anti-aliasing is especially important in \ac{DiDiGAN}'s generator. The anti-aliasing issue raised in StyleGAN3~\cite{stylegan3} translates to uncontrollable structural variations unrelated to the disease. Despite the use of a structural constrain $c$, the model is unable to maintain more fine-grained anatomical correspondence (especially in the cortical areas) due to aliasing. Example results synthesised without anti-aliasing can be found in Fig.~\ref{fig:s3}

\begin{figure*}[ht]
 \centering
 \includegraphics[width=0.8\textwidth]{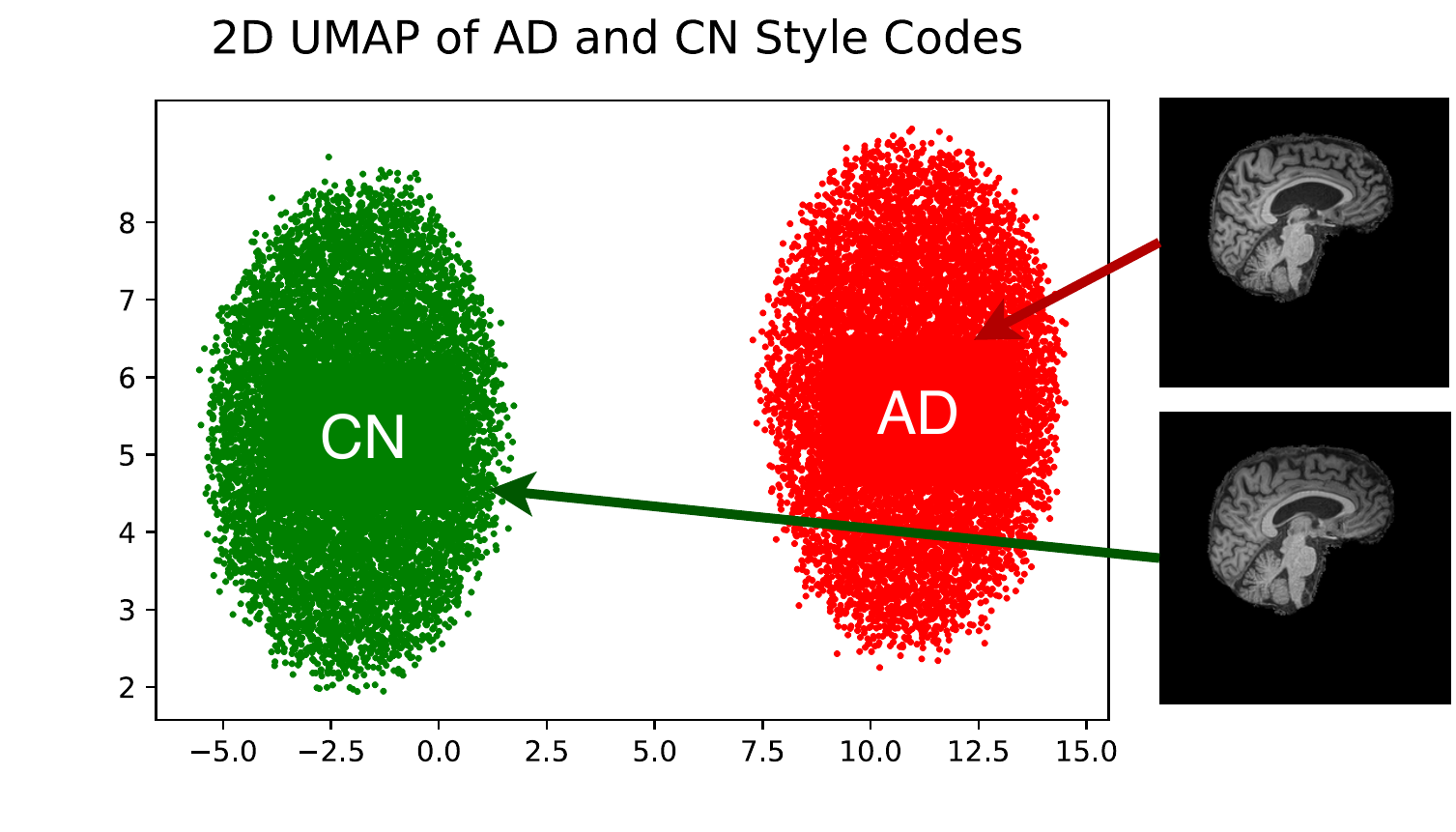}
 \caption{2D UMAP visualisation of the 512-dimensional manifold. 10,000 style codes from each domain are sampled and the UMAP algorithm is configured to focus on global structures in the distribution.}
 \label{fig:manifold}
\end{figure*}
Figure~\ref{fig:manifold} shows a 2D UMAP~\cite{umap} projection of \ac{AD} and \ac{CN} style codes sampled from the manifold (with different noise input). The two classes (disease / non-disease) form two distinctly separate clusters on the manifold. There is no overlap between the two distributions by design, which ensures \ac{AD} features can be observed from any \ac{AD}-\ac{CN} pair regardless of the noise vectors used. It was also noted that traversing the manifold through different noise vectors within each disease / non-disease  class introduces minor non-disease (such as contrast) variations.

\begin{table*}[ht]
    \small
    \centering
    \caption{2D and 3D \ac{AD} and \ac{CN} classification results on the ADNI dataset.}
    \begin{tabular}{llll}
    \textbf{Ref}             & \textbf{Method}                           & \textbf{Test Accuracy} & \textbf{2D/3D} \\
    \hline
    Valliani et al. \cite{valliani2017deep} & ResNet-18                        & 81.3\%  &2D \\
    Wen et al. \cite{wen2020convolutional}      & Pre-trained ImageNet + fine-tuning            & 79\%    &2D \\
    \hline
    Li et al. \cite{li2017alzheimer} & 3D CNN & 88\% & 3D\\
    Senanayake et al \cite{senanayake2018deep} & 3D CNN & 76\% & 3D\\
    Korolev et al \cite{korolev2017residual} & 3D CNN & 80\% & 3D\\
    Cheng et al \cite{cheng2017classification} & 3D CNN & 87\% & 3D\\
    \hline
    \textbf{\ac{DiDiGAN}}         & Pre-trained $D$ \textbf{without} fine-tuning & \textbf{82.3\%} & 2D\\
    \textbf{\ac{DiDiGAN}}            & Pre-trained $D$ + fine-tuning & \textbf{86.5\%} & 2D
    \end{tabular}
     \label{tab:class}
\end{table*}
A 2D U-Net~\cite{unet} was used to estimate (slice by slice) the systematic brain volume loss exhibited in the generated coronal \ac{AD} and \ac{CN} pairs. The U-Net was trained for 3-class segmentation (grey matter, white matter and \ac{CSF}) on the real data using labels generated from SPM~\cite{tzourio2002automated}) (because no manual ones are available). Although these labels are not manual, the quality would suffice for estimating a general trend for two large groups of samples. The fake samples to be analysed were generated using down-sampled real test data as constraints and $y_{ad}$ and $y_{cn}$ as disease conditions. When running inference on the fake data slice by slice, the U-Net did not output extreme values (such as excessively different area estimates). This is an indication that the generated data distribution is similar to the real one as deep learning tend to fail catastrophically on unseen distributions. Between the synthesised \ac{AD} and \ac{CN} slices, the U-Net predicted average shrinkage values of 5.5\%, 6.8\% and 10.4\% for the grey matter, white matter and \ac{CSF}, respectively, which is in line with the trend found in the real data.



\ac{DiDiGAN}'s ability to learn and synthesise \ac{AD} features can be further verified through 2D \ac{AD}-\ac{CN} classification (on the \ac{ADNI} dataset). It was noted that without any re-tuning, the $D_{w}$ head of a pre-trained discriminator already demonstrated good classification accuracy on the test set despite being limited to 2D. (Table~\ref{tab:class}). With some re-tuning (1000 samples), the discriminator easily exceeds the test set accuracy of other 2D comparable methods, even approaching the accuracies of some 3D methods~\footnote{A large number of classification works show varying degrees of data leakage and biases according to \cite{wen2020convolutional}. Hence only methods with no data leakage are cited as the baseline.}. Since the generator's performance is tightly coupled with the discriminator, a good \ac{AD} classification accuracy induces more confidence in the validity of the synthesised samples.

There are, however, some limitations of the current work. Since \ac{DiDiGAN} is a weakly-supervised framework, evaluation is challenging due to the lack of labels specific to disease appearances. Indirect and manual methods, as shown in this work, are still required in the process. Nonetheless, \ac{DiDiGAN}'s visualisation could still direct the focus of manual analysis to potential ROIs. As a \ac{GAN} framework, there is some inherent instabilities during the training process, manual monitoring of convergence is required due to the lack of universal quality metrics for \acp{GAN} in medical image analysis.

\ac{DiDiGAN} presents a range of directions for future work. First, the manifold clusters should be further investigated for insights. It would also be valuable to encode additional clinical information such as age and gender onto the manifold to assess relationships between different data formats. Second, the structural constraint could incorporate human-decision making in the loop to direct the ``attention" of \ac{DiDiGAN}. For example, selectively down-sampling human-defined ROIs while leaving the rest of the constraint in full resolution could capture disease characteristics local to those ROIs. Finally, \ac{DiDiGAN} showed promising results on the \ac{ADNI} dataset. It would be valuable to test it on datasets collected for other diseases.

\section{Conclusion}
\ac{DiDiGAN} demonstrated an excellent capacity to discover and aid disease characteristic visualisation from weakly-labelled medical images. Since many disease characteristics are extremely subtle,\ac{DiDiGAN} learns a disease manifold to create a pair of \ac{AD} and \ac{CN} images with a shared anatomical structure. The major technical novelties of \ac{DiDiGAN} are I) the use of a learnt disease manifold to influence the expression of \ac{AD} features and II) mechanisms including the structural constraint and III) anti-aliasing to maintain anatomical correspondence. In the experiments involving the ADNI dataset, \ac{DiDiGAN} discovered key \ac{AD} features, including reduced hippocampal volume, ventricular enlargement and cortical atrophy. These results demonstrated \ac{DiDiGAN}'s capability as a data-driven disease characteristic discovery method and future work will apply it to more dataset involving different diseases.

\bibliographystyle{IEEEtran}
\bibliography{ref}
\def\thesection{S\arabic{section}}

\newpage
~
\newpage
\setcounter{section}{0}
\setcounter{figure}{0}
\renewcommand{\figurename}{Fig.}
\renewcommand{\thefigure}{S\arabic{figure}}
\onecolumn
\section{Jacobian Index Computation}
2-dimensional non-linear Symmetric Diffeomorphic Image Registration with Cross-Correlation (SyN-CC) (Avants et al. 2008) cost function as implemented in Advanced Normalization Tool (ANTS v.2.3.4 (Kim et al. 2008)) was conducted with the generated CN images being the fixed images and the generated AD images being the moving images. The $antsRegistration$ command was used for this image registration step. Jacobian Indices were calculated from the warp field obtained from the image registration step above using the $CreateJacobianDeterminantImage$. The Jacobian Index map was scaled and displayed as log Jacobian, so that the distribution is symmetrical around zero, with positive and negative values indicating volume expansion and contraction, respectively.

\begin{figure*}[h]
 \centering
 \vspace*{2cm}
 \includegraphics[width=0.65\linewidth]{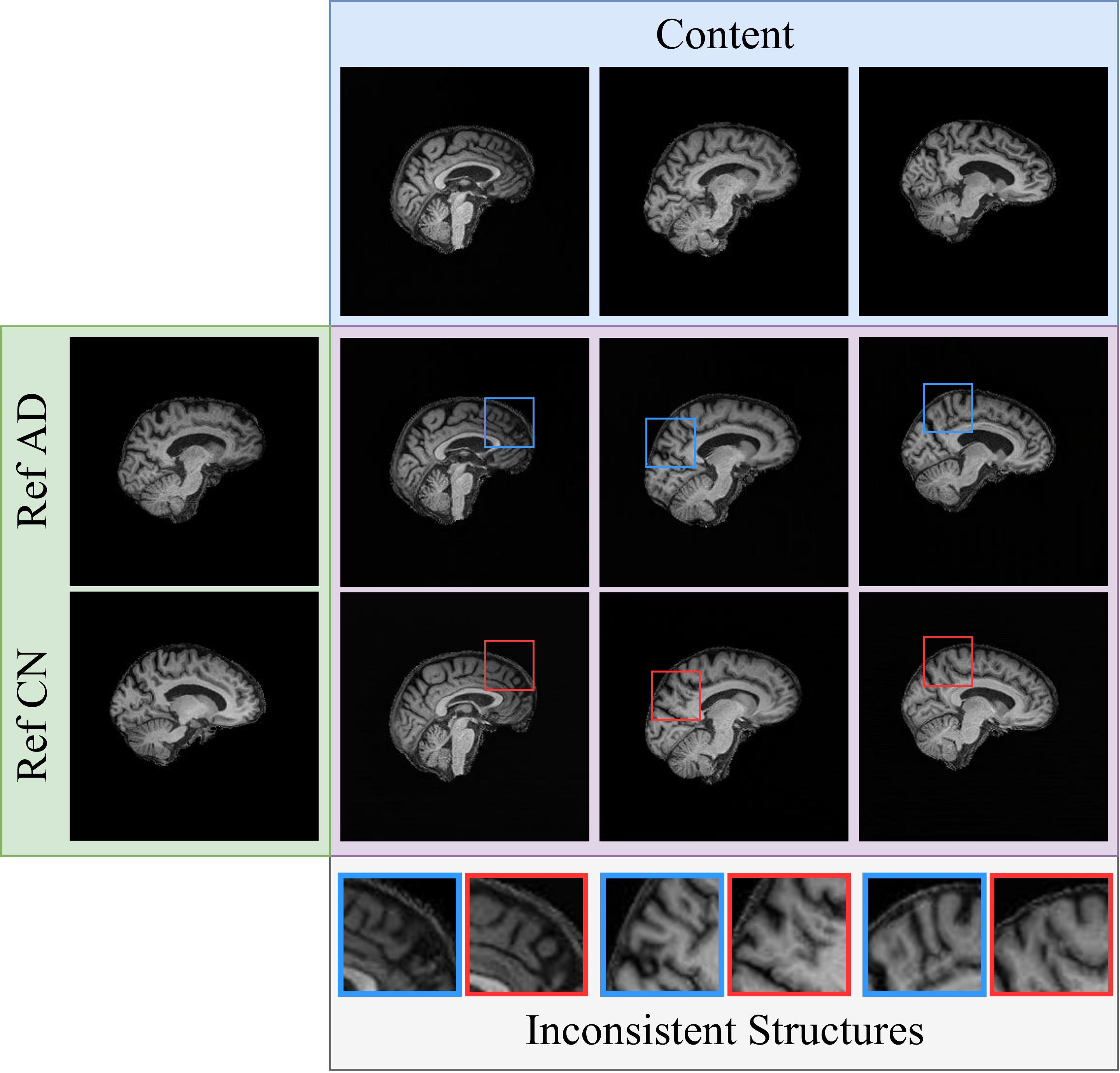}
 \caption{StarGANv2 failed to retain structural consistency for meaningful comparative analysis.}
 \label{fig:s1}
\end{figure*}

\begin{figure*}[h]
 \centering
 \includegraphics[width=0.65\linewidth]{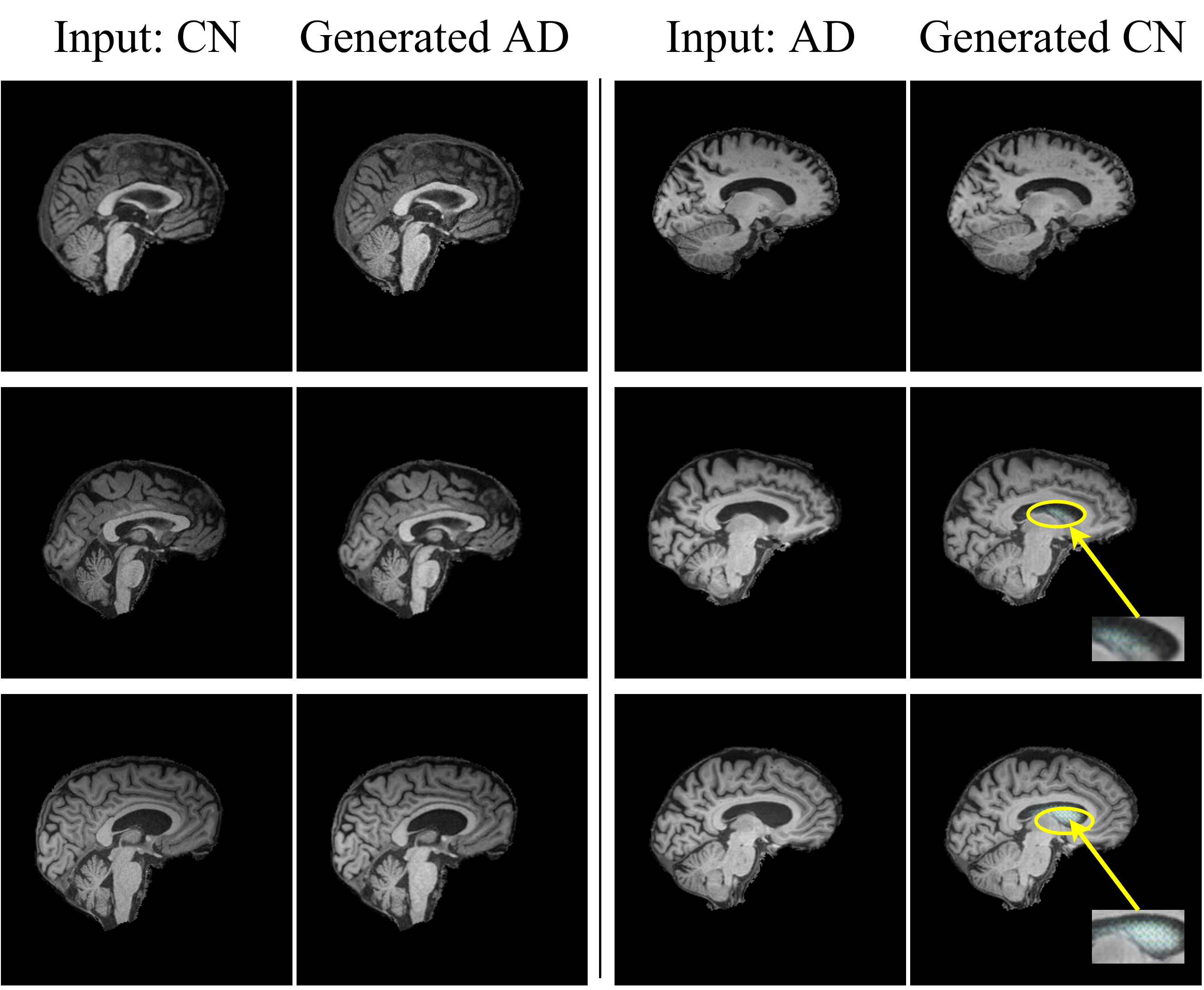}
 \caption{CycleGAN created almost identical \ac{AD} and \ac{CN} images and some result contained artifacts.}
 \label{fig:s2}
\end{figure*}

\begin{figure*}[h]
 \centering
 \includegraphics[width=1\linewidth]{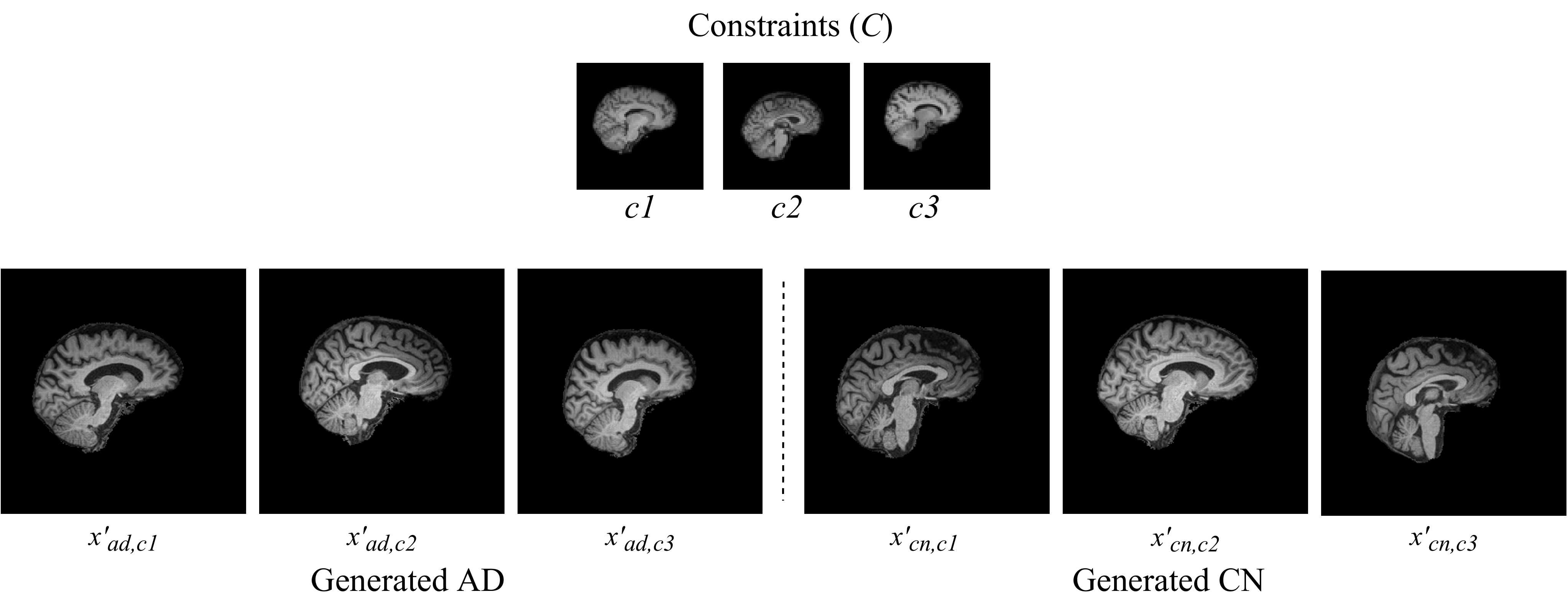}
 \caption{Results without anti-aliasing. $x'_{ad, c}$ and $x'_{cn, c}$ were generated from structural constraints $c1, c2$ and $c3$. There are excessive non-\ac{AD} variations which makes side-by-side comparison meaningless.}
\label{fig:s3}
\end{figure*}

\begin{figure*}[h]
 \centering
 \includegraphics[width=1\linewidth]{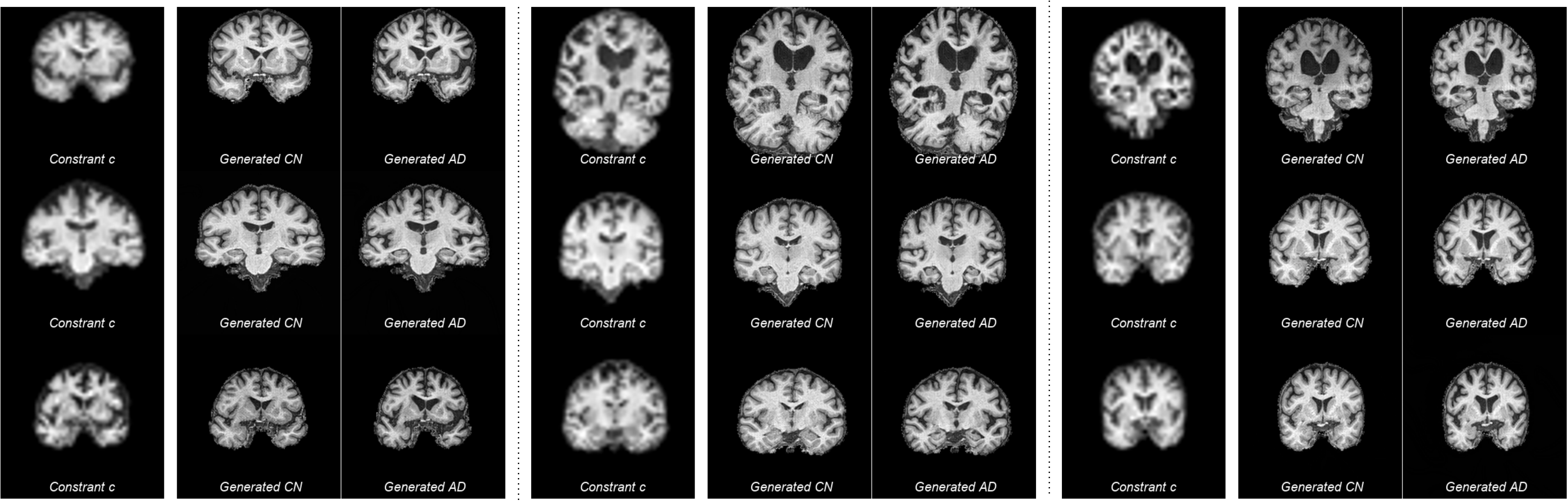}
 \caption{Additional results based on more diverse brain structures (more extreme slice locations).}
 \label{fig:s4}
\end{figure*}

\acrodef{GAN}{Generative Adversarial Network}
\acrodef{AD}{Alzheimer's Disease}
\acrodef{MRI}{magnetic resonance image}
\acrodef{CNN}{Convolutional Neural Network}
\acrodef{CN}{Cognitively Normal}
\acrodef{AdaIN}{Adaptive Instance Normalisation}
\acrodef{DiDiGAN}{Disease Discovery GAN}
\acrodef{DSC}{Dice Similarity Coefficient}
\acrodef{ADNI}{Alzheimer's Disease Neuroimaging Initiative}
\acrodef{CSF}{cerebrospinal fluid}

\end{document}